\documentclass[12pt]{article}
\usepackage[usenames,dvipsnames]{xcolor}
\usepackage{amsmath,amsfonts,amssymb}
\usepackage{graphicx}
\usepackage{psfrag}
\usepackage{enumerate}
\usepackage{mathrsfs}
\usepackage{dsfont}

\newcommand\blfootnote[1]{%
  \begingroup
  \renewcommand\thefootnote{}\footnote{\hspace{-6mm}#1}%
  \addtocounter{footnote}{-1}%
  \endgroup
}

\begin{document}
\clearpage\thispagestyle{empty}

\begin{center}


{\large \bf
Comments on Microcausality, Chaos, and Gravitational Observables}

%
%

\vspace{7mm}

Donald Marolf \\

\blfootnote{\tt marolf@physics.ucsb.edu}




\bigskip\centerline{\it Department of Physics, University of California,}
\smallskip\centerline{\it Santa Barbara, CA 93106, USA}


\end{center}

\vspace{5mm}

\begin{abstract}
Observables in gravitational systems must be non-local so as to be invariant under diffeomorphism gauge transformations.
But at the classical level some such observables can nevertheless satisfy an exact form of microcausality.  This property is conjectured to remain true at all orders in the semiclassical expansion, though with limitations at finite $\hbar$ or $\ell_{Planck}$.  We also discuss related issues concerning observables in black hole spacetimes and comment on the senses in which they do and do not experience the form of chaos identified by Shenker and Stanford.  In particular, in contrast to the situation in a reflecting cavity, this chaos does not afflict observables naturally associated with Hawking radiation for evaporating black holes.
\end{abstract}

\setcounter{footnote}{0}
\newpage
\clearpage
\setcounter{page}{1}

\tableofcontents

\section{Introduction} \label{sec:intro}

The algebra of gravitational observables has played important roles in several recent investigations of quantum gravity.  It was used  to analyze black hole scrambling and chaos in \cite{Shenker:2013pqa,Shenker:2013yza,Shenker:2014cwa,Maldacena:2015waa}.  It was discussed in connection with quantum error correction properties of a dual CFT in \cite{Almheiri:2014lwa}. And it was studied directly in \cite{Giddings:2015lla,Donnelly:2015hta} to gain insight into the structure of quantum gravity.  A common theme in the above works has been that transforming a bare local operator into a fully-dressed diffeomorphism-invariant observable can  significantly alter the associated commutators. In particular, the delocalized nature of this gravitational dressing can lead to violations of naive microcausality.  By this we mean that commutators between two operators can be nonzero when bits of their dressings are causally related, even though the bare parts of the operators remain spacelike separated.  See for example the explicit computations in \cite{Donnelly:2015hta}.   In contrast, we focus below on classes of observables for which taking the bare parts to be spacelike separated makes commutators either precisely zero or extremely small.  In this sense their algebra is much closer to the algebra of local operators in non-gravitating field theories.

We first review a class of relational observables
introduced long ago \cite{GD1,GD2,Komar1958,BergmanKomar1960,DeWitt:1962cg} and which exhibits exact microcausality at the classical level -- at least in suitably generic states for which infrared (IR) issues can be neglected.   We use the name single-integral observables following \cite{Giddings:2005id,Giddings:2007nu}.   Such observables utilize structures in the state to define a physical coordinate system analogous to Einstein's rods and clocks and with respect to which more-or-less localized quantities can then be defined.  In essence, these are the natural observables in a Higgsed phase of gravity where diffeomorphism-invariance is spontaneously broken and can be effectively ignored. See \cite{Khavkine:2015fwa} for a recent discussion of related observables and their classical locality properties.

Our main interest below concerns quantum versions of single-integral observables. We conjecture that the exact microcausality mentioned above continues to hold at all orders in the semiclassical expansion, though there are non-perturbative limitations at finite $\hbar$ or $\ell_{Planck}$.  Both classical and quantum versions are discussed in section \ref{sec:SI}.

Section \ref{sec:BH} then comments briefly on related implications for the physics of black holes. The main point is to note various ways in which naive analyses might suggest that large commutators from gravitational dressings play important roles in the black hole information problem\footnote{The arguments in \cite{Shenker:2013pqa,Shenker:2013yza,Shenker:2014cwa,Maldacena:2015waa} are quite distinct from those critiqued here.  Our remarks in no way diminish their implications for scrambling.}.  In each case we find that, by dressing the operators carefully, such commutators can be either drastically reduced or avoided altogether.  We close with some final discussion in section \ref{sec:Disc}.

\section{Single-Integral Diffeomorphism-invariant Observables}
\label{sec:SI}

Theories with massless gravitons treat many diffeomorphisms as guage symmetries.  Those that are not gauge act non-trivially at (perhaps asymptotic) timelike or null boundaries\footnote{Imposing conditions at spacelike boundaries restricts the possible initial data, typically resulting in a degenerate phase space.  And without special boundary conditions, spacelike boundaries act much the same as any other spacelike surface and do not break diffeomorphism invariance.  We therefore use the term boundary to refer only to timelike and null boundaries below, though this does not prohibit the spacetime having additional spacelike boundary components at singularites, at future or past infinity, or at otherwise-regular spacelike surfaces.}, though precisely which diffeomorphisms remain pure gauge can depend on details of the boundary conditions.  One may thus classify observables into two categories depending on whether they are in fact invariant under {\it all} diffeomorphisms of the spacetime, or only under those that are gauge.   The latter type effectively use the boundary as a fixed structure relative to which physics can be defined.  They may thus be called boundary-relational, while we reserve the term diffeomorphism-invariant for the first category above.  In contexts without timelike or null boundaries, all diffeomorphisms are gauge and all observables must be fully diffeomorphism-invariant. While it is often convenient to use quantum language below, our initial discussion will be strictly classical.  Quantum issues will be addressed separately in sections \ref{quantum} and \ref{IR}.

The recent works \cite{Shenker:2013pqa,Shenker:2013yza,Shenker:2014cwa,Almheiri:2014lwa,Giddings:2015lla,Maldacena:2015waa,Donnelly:2015hta} considered contexts with boundaries and focused on boundary-relational observables.  One class of examples is built from the associated boundary values of bulk fields or -- at an asymptotic anti-de Sitter (AdS) boundary --  the corresponding rescaled asymptotic values  that define local CFT operators under the AdS/CFT dictionary \cite{Witten:1998qj}.  Related observables inside the bulk may be constructed from e.g. the values of scalars at some given (perhaps renormalized) distance into the bulk along a geodesic fired orthogonally into the interior from a given boundary point.    Such constructions were argued  in \cite{Almheiri:2014lwa} to be analogous to Wilson line observables in Yang-Mills gauge theories and we follow \cite{Donnelly:2015hta} in refering to them as gravitational Wilson lines ( see also the related gravitational work \cite{Tsamis:1989yu}).
It was also noted in \cite{Almheiri:2014lwa} that such observables are equivalent to those defined by Fefferman-Graham gauge conditions \cite{FG} (thus making contact with \cite{Heemskerk:2012np,Kabat:2013wga,Kabat:2015swa}), and that this is again analogous to a kind of axial gauge for Yang-Mills fields.

Gravitational Wilson line observables transform non-trivially under any isometries of the boundary and so carry gravitational charges like energy,  momentum, and angular momentum. The gravitational Gauss law then requires them to have non-zero commutators with purely gravitational boundary observables whose integrals give the associated total charges.  One often says that such observables carry gravitational tails along the geodesics used in their definition, or one refers to this commutator as resulting from the gravitational dressing of the bare local (and thus not yet diffeomorphism-invariant) scalar.  Of course, there are many more possible types of gravitational tails and dressings with similar properties.  Some of these involve averaging the above construction over some family of geodesics \cite{Donnelly:2015hta}, but one can imagine many others as well.

The fact that boundary-relational observables carry gravitational charges gives them certain properties similar to familar local fields.  Acting with such observables -- i.e., moving along the flow they generate via the Poisson bracket\footnote{We use the term Poisson bracket in the sense in which it can be evaluated between two functions at different times, or even between functions non-local in time; see e.g. \cite{Marolf:1993zk}. For pedagogical purposes, it would be better to refer to the Peierls bracket \cite{Peierls:1952cb}, though this is less familiar to most readers.} -- changes the energy in a very physical way, and (asymptotic) translation and rotation symmetries can be used to move them from point to point in the bulk.  But the non-zero commutator with gravitational fields at the boundary constitutes a failure of microcausality, and thus a departure from local field theory expectations.  It is clearly a sign that the quantum gravity Hilbert space is less local than its quantum field theory counterpart, in the sense that there is even less factorization between spacelike separated regions.  This observation dates back to at least \cite{Lich} who noted at the classical level that Einstein's equations involved constraints forbidding independent choices of initial data even in spacelike separated regions, though see  e.g. \cite{Giddings:2001pt,Giddings:2004ud,Marolf:2008mf,Donnelly:2014gva,Donnelly:2014fua,
Giddings:2015lla,Donnelly:2015hxa,Bousso:2015mna} for more recent discussions explicitly emphasizing this point in the quantum context as well as e.g. \cite{Higgs:1958mh,DeWitt:1962cg,Ashtekar:1991hf,Torre:1992rg,Freidel:2005me} for closely related comments.

To correctly interpret the implications of this observation, it is important to understand whether other constructions of gravitational observables might better preserve microcausality.  The above discussion makes clear that it would be best for such observables to carry no gravitational charges\footnote{One may say that the gravitational dressing cancels the would-be bare gravitational charges.  In other words, our observables redistribute energy, momentum, etc. between various types of excitations.  As a result, such observables act trivially on any unique minimum-energy vacuum state.}.  They are thus invariant under all diffeomorphisms that preserve the given boundary conditions.  Extrapolating this result, it is natural to study those that are fully diffeomorphism-invariant.

We will focus on a particular class of such observables introduced by Geheniau and Debever \cite{GD1,GD2} (see also \cite{Komar1958,BergmanKomar1960}).  In $d$ spacetime dimensions, one may take any $d$ independent scalars $Z^\alpha$ to act as coordinates, at least in sufficiently local regions.  We require our $Z^\alpha$ to be local, in the sense that they are locally constructed from finitely many derivatives of fields with dynamics governed by local actions that yield causal equations of motion.

Given a $(d+1)$th local scalar $\phi$ and $d$ numbers $Z_0^\alpha$, one may consider the values of $\phi$ at  spacetime points where the $Z^\alpha$ take the specified values $Z_0^\alpha$.  When there is a unique such point, the corresponding $\phi$-value defines a diffeomorphism-invariant observable $[\phi](Z_0)$.  More generally, it is useful to follow the construction given by DeWitt \cite{DeWitt:1962cg} and use
\begin{equation}
\label{eq:SingInt}
[\phi](Z_0) = \int d^dx \ \phi \  \delta^{(d)}(Z^\alpha - Z_0^\alpha) \
|\frac{\partial Z}{\partial x}|.
\end{equation}
This formulation gives a finite value for $[\phi](Z_0)$ on any solution where the condition $Z^\alpha = Z_0^\alpha$ is satisfied at a finite number of points, including the case where it fails to be satisfied anywhere (in which case $[\phi](Z_0)=0$).   As we discuss further in section \ref{IR}, the integral in \eqref{eq:SingInt} can diverge when an infinite number of points satisfy $Z^\alpha = Z_0^\alpha$.
Similar constructions for general tensors, spinors, etc. are clearly possible as well \cite{DeWitt:1962cg}, though for simplicity we restrict ourselves to scalars.  We mention in passing that there are many other possible constructions (see e.g. \cite{Mukhanov:1990me,Rovelli:2001my,Khavkine:2011kj,Duch:2014hfa}) of observables that may be seen as hybrids of the Geheniau-Debever construction with gravitational Wilson lines and which have varying degrees of microlocality; another class of such examples will be further discussed in section \ref{IR}.

With an eye toward the quantum problem, it is natural to follow  and further generalize \eqref{eq:SingInt} to allow the
integrand to be a smooth function on the space of fields.  In particular, we might consider observables of the form
\begin{equation}
\label{eq:SingIntSmeared}
[\phi](Z_0) = \int d^dx \ \phi \ f(Z^\alpha - Z_0^\alpha) \
\det\left( \frac{\partial Z}{\partial x} \right)
\end{equation}
for general smooth functions $f$.  In this work we will take $f$ to have compact support.    The resulting \eqref{eq:SingIntSmeared} may then be called compactly-supported single-integral observables, though one should realize that this refers to compact support in $Z$-space and not necessarily in the physical spacetime. The replacement of the delta-function from \ref{eq:SingInt} by the smooth function $f$ was previously discussed in \cite{Giddings:2005id,Gary:2006mw,Giddings:2007nu}, and in particular \cite{Gary:2006mw} provides an explicit example of a similar construction in two dimensions. Dropping the absolute value on the Jacobian was discussed in \cite{Marolf:1994wh,Marolf:1994nz}. This latter change makes no difference in perturbation theory around a background where the Jacobian is everywhere non-vanishing, but could be useful at the non-perturbative level\footnote{In particular, one might expect that typical field eigenstates satisfy $Z^\alpha - Z_0^\alpha$ at an infinite number of spacetime points.  For positive operators $\phi$, keeping the absolute value would then lead to an infinite set of positive contributions.  But choosing instead the form \eqref{eq:SingIntSmeared} would allow cancellations.}.

At the classical level, the observables \eqref{eq:SingInt} and \eqref{eq:SingIntSmeared} satisfy the following exact version of microcausality \cite{DeWitt:1962cg}.  Recall that the Poisson Bracket $\{A, B\}$ of observables $A, B$ is a function on the gravitational phase space, or equivalently on the space of solutions.  For $A,B$ of the form \eqref{eq:SingInt} associated respectively with $Z_0^\alpha$-values $Z_A^\alpha$, $Z_B^\alpha$, this bracket vanishes when evaluated on solutions for which all spacetime points with $Z^\alpha = Z_A^\alpha$ are spacelike separated from all those having $Z^\alpha = Z_B^\alpha$.  For $A,B$ of the form \eqref{eq:SingIntSmeared} defined by compactly supported functions $f_A,f_B$, the bracket vanishes on solutions for which the spacetime supports of $f_A(Z^\alpha - Z_A^\alpha)$ and $f_B(Z^\alpha - Z_B^\alpha)$ are spacelike separated.

There are many simple ways to see this microcausality.  Perhaps the most direct is to summarize and paraphrase the argument from \cite{DeWitt:1962cg} based on the Peierls Bracket.  The Peierls bracket \cite{Peierls:1952cb} is a more covariant structure equivalent to the Poisson bracket in the sense defined here (and when acting on gauge-invariant quantities) but which can be built directly from advanced and retarded Green's functions for the linearized equations of motion; the linearization is performed about the solution ${\cal S}$ on which the bracket is to be evaluated.  In particular, $\{A, B\}$ is the difference between the linearized advanced and retarded changes in $A$ when $B$ is added as a source to the action.  The microcausality is an immediate consequence of the vanishing of the above Green's functions vanish at spacelike separations.

For later purposes, it is useful to repeat the above derivation using a slightly more pedestrian approach. Recall that the Poisson bracket is a derivation, meaning that for functionals $A[\psi], B[\psi]$ of fundamental fields $\psi$, the bracket $\{A[\psi], B[\psi]\}$ can be computed in terms of functional derivatives $\frac{\delta A}{\delta \psi}$, $\frac{\delta B}{\delta \psi}$ of $A, B$ and Poisson Brackets of $\psi(x)$ with $\psi(x')$.  Since $\psi(x)$ is not an observable, its Poisson brackets generally depend on a choice of gauge\footnote{Some readers will naturally interpret the term Poisson bracket in the sense used by Dirac \cite{Dirac:1964LQM}.  But any gauge fixing scheme naturally defines its own Poisson bracket which can be extended to act on non-local functions of time.  Indeed, as discussed in e.g. \cite{Marolf:1993zk,Marolf:1993af} a complete gauge-fixing is unnecessary.  And in this sense Dirac's formalism \cite{Dirac:1964LQM} for gravity is effectively just a choice of gauge associated with fixing lapse and shift.}, though any such choice must lead to the same final result for $\{A[\psi], B[\psi]\}$.

Consider then a solution ${\cal S}$ on which both functions $f_A(Z^\alpha - Z_A^\alpha), f_B(Z^\alpha - Z_B^\alpha)$ have compact support in spacetime.  When evaluated on ${\cal S}$, the functional derivatives $\frac{\delta A}{\delta \psi(x)}$ vanish at points $x$ outside the support of $f_A(Z^\alpha - Z_A^\alpha)$, and similarly for $B$.  Using a covariant gauge in which Poisson brackets of $\psi(x)$ vanish outside the light cone then establishes the desired result.  Note that locality of the $Z^\alpha$ and $\phi$ played a key role in this argument.  While one may also use \eqref{eq:SingInt}, \eqref{eq:SingIntSmeared} to define observables built from non-local scalars $Z^\alpha, \phi$, they would no longer obey simple microcausality relations.

\subsection{Quantum Microcausality?}
\label{quantum}

The above language and reasoning are both very classical.  At the quantum level the evaluation of $\{A, B\}$ on a solution ${\cal S}$ is replaced by taking the expectation value of $[A,B]$ in some state, or perhaps by asking if $[A,B]$ annihilates the given state.  But it seems unlikely that quantum states at finite $\ell_{Planck}$ admit precise notions of spacelike separated regions.  There are also non-perturbative infra-red issues that will be discussed in section \ref{IR} below\footnote{See also footnote \ref{flucts} for comments on ultraviolet (UV) issues involving fluctuations.}.  In this sense we expect microcausality to be at best a semiclassical phenomenon.

It nevertheless remains very interesting to investigate just how quickly microcausality emerges as $\ell_{Planck} \rightarrow 0$.  We quantify this by asking how small the corrections can be made in an appropriate semiclassical limit. Here we expand all fields $\psi = \psi_{cl} + \delta \psi$ in terms of the quantum fluctuation $\delta \psi$ around a classical background $\psi_{cl}$ given by evaluating the fields fields $\psi$ on some classical solution ${\cal S}$. The background $\psi_{cl}$ may include gravitational back-reaction determined by some finite Newton constant $G$, so an expansion in $\ell_{Planck}$ is equivalent to the standard semi-classical expansion in powers of $\hbar$, and thus in powers of the quantum fluctuations $\delta \psi$. We consider observables $A, B$ of the form \eqref{eq:SingIntSmeared} for which the supports of $f_A(Z^\alpha - Z_A^\alpha), f_B(Z^\alpha - Z_B^\alpha)$ are spacelike separated in ${\cal S}$.

At first order in $\delta \psi$, the observables $A,B$ are characterized by the first order changes $\delta A$, $\delta B$ around their background values.  These $\delta A$, $\delta B$ are linear combinations of the $\delta \psi$, weighted by the relevant functional derivatives of $A,B$.  As noted above, these functional derivatives vanish outside the support in ${\cal S}$ of $f_A(Z^\alpha - Z_A^\alpha), f_B(Z^\alpha - Z_B^\alpha)$.  So using a covariant gauge to compute commutators of $\delta \psi$ again gives exact microcausality.

Indeed, with natural choices the above argument can be repeated at all orders in the semiclassical expansion.  An any finite such order we replace $A, B$ by renormalized polynomials in $\delta \psi$.  Due to factor ordering ambiguities, these polynomials are not fully determined by the classical Taylor series for $A, B$.  But it is nevertheless natural to take the coefficients to vanish outside the region in which $f_A(Z^\alpha - Z_A^\alpha), f_B(Z^\alpha - Z_B^\alpha)$ are supported\footnote{\label{Weyl}This may be achieved, for example, by defining the integrands via a Fourier transform in field space.  For example, consider the expression $f(Z-Z_A) = \int dk e^{ik_\alpha (Z^\alpha-Z^\alpha_0)} \tilde f(k)$. Expanding the exponential to a finite order about a classical background and integrating gives coefficients that are just derivatives of $f$ evaluated on the background as we desire. A Fourier transform representation of this sort was used in \cite{Giddings:2005id,Gary:2006mw} to study a Weyl-invariant two-dimensional model, though the so-called gravitational dressing factors used there are non-polynomial in $k^\alpha$ even in perturbation theory and would thus spoil this argument.  The arguments here indicate that such dressing factors can be avoided when one can choose a preferred physical metric for use in renormalizing the operators; i.e., at least for Einstein-Hilbert-like theories of gravity in which Weyl rescalings are not gauge transformations.}.  Continuing to work in a covariant gauge one then finds exact microcausality at each order in $\delta \psi$. Note that at each such order our observable effectively reduce to those studied in \cite{Khavkine:2015fwa}.

One may ask whether there are important effects from the renormalizations that replace singular products of fundamental fields with better-defined composite operators.  But in a covariant gauge this process proceeds much as in a quantum field theory without gauge symmetry.  When covariant techniques are used as in \cite{Brunetti:2001dx,Hollands:2014eia}, it manifestly preserves microcausality.  It also preserves the tensor (or tensor-density) character of the desired operator, so that the renormalized polynomial defined by the integrand of \eqref{eq:SingIntSmeared} is indeed a density in the sense of \cite{Brunetti:2001dx,Hollands:2014eia}. In this sense the integrals of such quantities, and hence our observables $A,B$, remain invariant under diffeomorphisms. This program was recently implemented for perturbative gravity in \cite{Brunetti:2013maa}.

However, a question remains as to whether the quantum observables $A, B$ -- defined by the above power series expansions in $\delta \psi$ -- are truly independent of the choice of gauge.  For example, do the same power series evaluated in another gauge, perhaps a different covariant gauge or a perhaps a non-covariant gauge,  continue to satisfy the same algebra?  For Yang-Mills theories it is known \cite{Barnich:1994ve,Hollands:2007zg} (with the latter building on \cite{PS,Duetsch:2001sw,Brennecke:2007uj}) that any local quantity that is gauge-invariant at the classical level can be taken to define a BRST-invariant quantum operator, demonstrating full gauge-invariance at the quantum level.     While there appears to be no corresponding result in the literature for perturbative gravity, in the absence of known gravitational anomalies it is natural to conjecture\footnote{Though some opinions differ.  I thank Tom Banks for discussions on this issue.} that an analogous construction exists allowing series of the above form to define appropriately gauge-invariant quantum operators at each order in $\ell_{Planck}$, and while simultaneously retaining the exact microcausality discussed above. We note that an appropriate notion of BRST charge was recently constructed by Brunetti et al \cite{Brunetti:2013maa} using the desired covariant techniques, so it remains only to address the invariance of single-integral observables.  We leave this for future work.

\subsection{Infrared Issues}
\label{IR}

We close this section with a brief discussion of infra-red issues, especially at the non-perturbative level.  The details may well be interesting to explore in the future, but we content ourselves here with certain general remarks.

At the classical level, the convergence of \eqref{eq:SingInt} on a solution ${\cal S}$ is guaranteed if the condition $Z^\alpha =Z_0^\alpha$ is satisfied in ${\cal S}$ at only finitely many spacetime points.
The same is largely true of \eqref{eq:SingIntSmeared}.
One generally expects this condition to hold in both asymptotically flat solutions and asympotically de Sitter solutions, and also for asymptotically anti-de Sitter solutions when the latter contains a black hole.   In all of these situations fields naturally decay to a background in which we can take \eqref{eq:SingIntSmeared} to vanish, so non-zero contributions should come from only an appropriately-finite region of spacetime.

In contrast, horizon-free asymptotically anti-de Sitter solutions tend to be quasi-periodic, and so can lead to IR divergences.  In order to be useful in this context, the observables  \eqref{eq:SingInt} and \eqref{eq:SingIntSmeared} require modifications.  The natural choice is to add just enough boundary-relational ingredients to tame these IR issues.  For example, one might insert factors into the integrand that vanish outside some time interval $[t_1,t_2]$ defined relative to the boundary.  Such factors spoil the argument for microcausality but, since Poisson Brackets involve derivatives, only by a small amount when $\Delta t = t_1 -t_2$ is large.  I.e., while the Poisson Bracket of two naively-spacelike-related observables $A,B$ may fail to vanish over large regions of $Z_A$, $Z_B$ it will nevertheless be small for given $Z_A$, $Z_B$.

Were the goal only to obtain finite observables, we could in fact choose $\Delta t$ to be arbitrarily large.  This would make the microcausality-violating term arbitrarily small in the commutator of any two given observables.  But such observables would effectively integrate over many semiclassical recurrences.  And since recurrences are widely separated in time, with AdS asymptotics it is difficult -- if not impossible -- to construct two such observables $A, B$ where all recurrences sampled by the first are spacelike separated from all recurrences sampled by the second.  There may thus be no values of the associated $Z_A, Z_B$ where $[A,B]$ is small.

Instead, microcausal behavior will be manifest only when the boundary-relational ingredients select a spacetime volume small enough that recurrences are unlikely.  Thus $\Delta t$ should be small compared to the timescale for quasi-periodic behavior.  The point, however, it that it can still remain large with respect to the timescale on which one hopes to localize the observable; the remainder of the localization can be done by the local fields $Z^\alpha$ without introducing further violations of microcausality.

Let us now consider the quantum version of such recurrence effects.  For simplicity we begin by addressing contexts with time-translation invariance and a well-defined Hamiltonian $H$.  Gravitational systems satisfying the most familiar asymptotically flat or asymptotically anti-de Sitter boundary conditions are of this form.  Then while the detailed behavior of operators like \eqref{eq:SingIntSmeared} in any particular state may be difficult to evaluate, on general grounds the timescale of quasi-periodic behavior is controlled by the density of states. In particular, we use the assumption that $H$ is a well-defined generator of time-evolution to avoid any need to directly discuss the features of the full quantum theory associated with any singularities that arise in the semiclassical time evolution.

Our discussion naturally breaks into two cases.  First consider situations in which
the spectrum of $H$ is continuous and the density of states is infinite.  For example, one expects this to be the case for asymptotically flat spacetimes.  The infinite density of states means that there are no quantum recurrences, so the observables \eqref{eq:SingIntSmeared} can generally be expected to have finite expectation values.

The second case arises when the spectrum of $H$ is discrete.  For definiteness, let us consider AdS/CFT for CFTs on a compact space cross time.  As in our classical discussion, the result depends critically on whether we study states whose classical limits describe black holes. Those that do are expected to have exponentially small level spacing $\sim e^{-S_{BH}}$, where $S_{BH}$ is the Bekenstein-Hawking entropy of the black hole. In contrast, at least at weak coupling the spacing of states that do not describe semiclassical black holes tends to be roughly $1/\ell$, where $\ell$ is AdS scale.

These results set the expected timescales over which the observables \eqref{eq:SingIntSmeared} must be modified in order for their integrands to remain well-localized.  So they must also determine the size of any commutator terms that violate microcausality as a result of such modifications. We thus expect to be able to define observables whose commutators violate microcausality by exponentially small amounts, at least in appropriate black hole states.

Let us briefly think through an enlightening physical example.  We can easily arrange a black hole spacetime in which the $Z$-fields are excited near $t=0$, perhaps such that $Z^\alpha = Z_0^\alpha$ at some finite collection of points.  As we evolve away from $t=0$, such excitations fall into the black hole and the fields take values far from $Z_0^\alpha$.  For a long time the spacetime becomes quite empty except for the black hole and some Hawking radiation.  Only after a recurrence time is there significant amplitude for configurations resembling $Z^\alpha = Z_0^\alpha$ to reappear.  So with moderate boundary-relational ingredients we can design observables that probe features near $t=0$ and display good approximations of microlocal behavior.  However, this example points out the large regions of very empty spacetime that arise between recurrences.  In such regimes one expects few observables with truly microlocal behavior.

Having addressed the AdS context, we should also explicitly discuss the asymptotically de Sitter (dS) setting.  As discussed in \cite{Giddings:2007nu}, the observables \eqref{eq:SingIntSmeared} are finite at the semiclassical level (at which there are no recurrences).  At the nonperturbative quantum level the situation is less clear, as the finite entropy of de Sitter space may be taken as an indication that the actual density of states is finite.  However, this may depend on the details of how the quantum Hilbert space is defined \cite{Willy,Banks:2000fe,Giddings:2007nu}.   We will simply follow e.g. \cite{Giddings:2007nu} in assuming that there are interesting quantum Hilbert spaces of (say, future-) asymptotically de Sitter spacetimes in which the actual density of states remains infinite and any issue of recurrences can be ignored. Other cosmological settings may also be interesting to consider, but it is generally difficult to address quantum recurrences without a solid understanding of how and to what extent classical singularities are resolved.

In addition to the above issues, it can happen that matrix elements of an operator ${\cal O}$ of the form \eqref{eq:SingIntSmeared} give convergent integrals in some basis of states, but that fluctuations of these operators diverge \cite{Giddings:2005id,Giddings:2007nu}.  As discussed in \cite{Giddings:2007nu}, this occurs when an infinite set of intermediate states contribute at sufficiently high levels to the computation of $\langle {\cal O}^2 \rangle$.  For example, in asymptotically de Sitter spaces one finds such an effect event at the semiclassical level, with the result that $\langle {\cal O}^2 \rangle$ is proportional to the volume of spacetime.   This will again entail modifications (perhaps along the lines described in \cite{Giddings:2007nu}) that likely violate microcausality.   However, the coefficient of the volume divergence is set by the probability per unit volume of finding a configuration with $Z^\alpha = Z_0^\alpha$ in the de Sitter vacuum\footnote{\label{flucts} Due to positivity of the conserved energy, this probability vanishes in asymptotically flat or asymptotically AdS spacetimes.  So in such cases the fluctuations are IR finite and require no further modifications of our observables.  This observation may also help to ameliorate non-perturbative UV issues associated with the fact that the integrand of \eqref{eq:SingIntSmeared} generally fails to a good operator-valued distribution when it is non-polynomial.}.  This can be quite small when $Z^\alpha$ is far from the vacuum expectation value of $Z^\alpha$.  So one again expects to need microcausality-violating ingredients only on large scales,   allowing the coefficient of the microcausality-violating term in the commutator to remain small\footnote{The exponential expansion of de Sitter space means that exponentially large volumes can arise from times that are only polynomially long. So in this case we expect the coefficient to be only polynomially small.}.

\section{Black Hole Spacetimes}
\label{sec:BH}

We now make three brief additional remarks about commutators of gravitational observables in states describing black holes.  The first two concern single-integral observables like \eqref{eq:SingIntSmeared}, while the last concerns boundary-relational observables relevant to evaporating black holes.

\begin{figure}
\begin{center}

\includegraphics[width=0.30\linewidth]{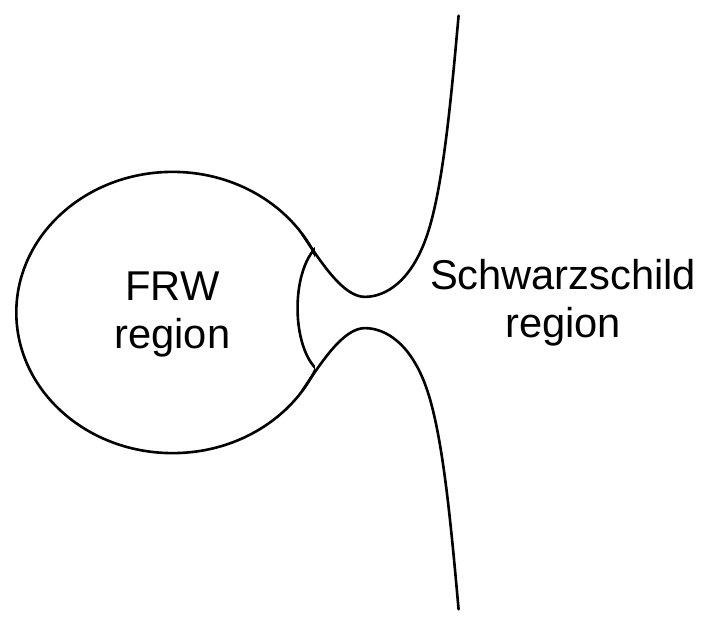}
 \caption{Moment of time-symmetry in a bag of gold spacetime constructed
by taking the two-sided Kruskal extension of Schwarzschild and replacing the 2nd asymptotic region with a spatially-compact Friedman-Robertson-Walker (FRW) cosmology.
}
\label{gold}
\end{center}
\end{figure}

\noindent
{\bf Counting Independent Observables Inside Black Holes:}  The degrees of freedom of classical black hole interiors are infinite in the UV (as in any region of space) and also in the IR.  The latter effect is associated with the infinite volume that can arise either from so-called bags of gold\footnote{This terms is often used to refer to eternal black holes with large-but-finite regions behind both future- and past event horizons, perhaps constructed as shown in figure \ref{gold}.  The actual connection to
\cite{Wheeler} is somewhat subtle, however.  The term was used differently in \cite{Wheeler}, though the spacetimes now called bags of gold were introduced there as well.  Also, in considering geometries like that in figure \ref{gold}, \cite{Wheeler} emphasized the physics on the spatially-compact cosmology side of the
Einstein-Rosen bridge as opposed to considering it as the
``inside'' of a black hole as we do here.} \cite{Wheeler} as in figure \ref{gold}, the related ``monster" constructions \cite{Hsu:2008yi,Hsu:2009kv}, or from evolution to late times (see e.g. \cite{Stanford:2014jda,Christodoulou:2014yia} for recent discussions of this time evolution).  This is in striking contrast with the black hole's finite Bekenstein-Hawking entropy $S_{BH} =A/4G$.

\medskip

Focusing on boundary-relational observables as in \cite{Shenker:2013pqa,Shenker:2013yza,Shenker:2014cwa,Maldacena:2015waa,Almheiri:2014lwa,Giddings:2015lla,Donnelly:2015hta} may seem to ameliorate this tension as it leads to the impression that the number of independent observables inside a black hole is proportional to its area.  In particular, if we wish to construct many Wilson-line type observables whose commutators satisfy microcausality, we must take care that no points on any Wilson line are causally related to those on any other -- effectively confining the entire set to a common spacetlike surface.  Assuming some minimal transverse smearing of each Wilson line then bounds the number of possible Wilson lines by a constant  -- presumedly of Planck-scale --  times the black hole area.

However, our diffeomorphism-invariant observables make clear that this supposed resolution was too quick.  Consider for concreteness a bag-of-gold construction where the cosmology on the far side of the Einstein-Rosen bridge has $Z$-field excitations near $t=0$ which happen to be arranged to provide good coordinates over large regions of spacetime.   Since there are infnite-volume cosmologies with finite energy density, one may have an arbitrarily large number of independent observables of the form \eqref{eq:SingIntSmeared} inside the black hole for which all pairwise commutators are microlocal to excellent approximation.

\medskip

\noindent
{\bf Commutators, Shock Waves, and Chaos:} The reader may
find it enlightening to discuss how the microcausality of \eqref{eq:SingInt} and \eqref{eq:SingIntSmeared} interacts with the chaos discussed in \cite{Shenker:2013pqa,Shenker:2013yza,Shenker:2014cwa,Maldacena:2015waa}.  We first note that one may argue for our microcausality using the techniques of \cite{Shenker:2013pqa}, which in their context gave exponentially large commutators for boundary-relational observables.  The key point is to consider the flows generated on phase space by any two such observables.  One first flows some finite distance along the Hamiltonian vector field of the first observable, and then along that of the second.  One compares the result with what one obtains when the order is reversed.  This gives the classical analogue of the commutator
\begin{equation}
e^{i\lambda_A A} e^{i\lambda_B B } -
e^{i\lambda_B B} e^{i\lambda_A A},  \ \ \  {\rm or \ perhaps \ of} \ \ \ e^{i\lambda_A A} e^{i\lambda_B B } e^{-i\lambda_A A} e^{-i\lambda_B B} ,
\end{equation}
of exponentials of the observables $A,B$.  Suppose that our observables are of the form \eqref{eq:SingInt} for $Z_0$-values given respectively by $Z_A^\alpha, Z_B^\alpha$, and let these observables act on a classical solution ${\cal S}$ where each set of values occurs at only one spacetime point, $p_A$ or $p_B$.  Since $A$ is non-trivial on ${\cal S}$ only at $p_A$ , the action of
$e^{i\lambda_A A}$ is easily described by considering a Cauchy surface for ${\cal S}$ through $p_A$. This action alters the corresponding Cauchy data by introducing at $p_A$ a disturbance in the $Z^\alpha$ and $\phi$ fields.   The new solution is then given by evolving the new Cauchy data both forward and backward in time.  The action of $e^{i\lambda_B B}$ may be treated similarly.

So long as $p_A,p_B$ are spacelike separated, we can choose to use a common Cauchy surface through both points to compute the effects generated by both
 $e^{i\lambda_A A}$ and $e^{i\lambda_B B}$.    It is then clear that modifying Cauchy data near $p_A$ commutes with modifying the data near $p_B$.  It is only when $p_A,p_B$ are causally related that the commutator can be nonzero.  This is the desired microcausality.  And at this purely classical level of analysis, the commutator continues to vanish (and thus fails to become exponentially large) even when the observables are separated by large times -- so long as their separation in space remains even greater, with at least one of the two being taken sufficiently far inside the black hole.  The story is similar for the observables \eqref{eq:SingIntSmeared}.

However, as discussed in section \ref{IR},  at the quantum level we often expect recurrences to force modifications of our observables. The addition of boundary-relational ingredients will then induce small failures of microcausality which grow exponentially in time due to the effects discussed in \cite{Shenker:2013pqa}.  The net effect is that the commutator of the modified observables is smaller than that discussed in \cite{Shenker:2013pqa,Shenker:2013yza,Shenker:2014cwa,Maldacena:2015waa} by an overall factor, though it grows at the same exponential rate.  As also discussed in section \ref{IR}, in favorable circumstances this overall factor might be exponentially small.  The commutator would then appear to remain small until close to the classical Poincar\'e (single-exponential) recurrence time\footnote{The time at which a given classical trajectory returns to near its initial location in phase space.}.

\medskip

\noindent
{\bf No chaos in Hawking evaporation:} As a side comment, we take the opportunity to mention that -- at least in the context of Hawking radiation -- the the chaos of
\cite{Shenker:2013pqa,Shenker:2013yza,Shenker:2014cwa,Maldacena:2015waa} is associated with black holes in reflecting (perhaps AdS) cavities.  In contrast, there is no such chaos for observables naturally associated with with Hawking radiation from evaporating black holes.

Consider for example an asymptotically flat black hole, where we  define observables $A,B$ associated with the Hawking radiation using  Wilson lines (perhaps suitably averaged over rotations in analogy with \cite{Donnelly:2015hta}) along half-infinite segments $\gamma_A, \gamma_B$ of past-directed null geodesics launched inward from widely separated points on future null infinity.  As in the above references, we may take the action of such observables $A,B$ to generate shock waves.  We compute their effects by modifying Cauchy data along the indicated Wilson lines and evolving to the past and future as above.  Since we are interested in outgoing radiation, we take $\gamma_A, \gamma_B$ to end far from the black hole.

When evolved to either the future or past, the shockwaves propagate along null geodesics.   So long as the geodesics widely separated, the interactions between them are weak.  In particular, since both $\gamma_A, \gamma_B$ end at large radius, the interactions are perturbatively small in the region lying both to the future of $\gamma_A$ and to the past of $\gamma_B$, or vice versa.   This suffices to make the commutator small in the sense of \cite{Shenker:2013pqa}. A similar argument again gives small commutators when an AdS black hole evaporates either due to the AdS scale $\ell$ being much larger than the black hole size, or due to  the AdS system being coupled to a large-entropy external system as in section 4 of \cite{Almheiri:2013hfa}.
Since there is no exponential growth  when $\gamma_A, \gamma_B$ are separated by many light-crossing times, such observables do not experience the chaos found in \cite{Shenker:2013pqa}.

\section{Discussion}
\label{sec:Disc}

This note has stressed the existence of diffeomorphism-invariant gravitational observables whose classical commutators respect a precise notion of microcausality.  Such observables can be constructed in sufficiently inhomogeneous regions of spacetime, where local quantities can be used to define an effective reference system.  The desired local quantities might be scalars formed from the Riemann tensor \cite{Komar1958} or its derivatives, or they might refer to local matter fields.  In such cases one can think of the theory -- at least within the given spacetime region -- as being in a Higgsed phase in which diffeomorphism-invariance is spontaneously broken.   At the classical level one finds such inhomogeneity in any sufficiently generic solution and the theory is effectively always Higgsed.   This provides a sharp sense in which microcausality is exact in the classical theory, and which may be considered an alternate formulation of the classical microcausality discussed in \cite{Khavkine:2015fwa}.

Our main point was to suggest that a form of microcausality survives quantum effects and remains exact at all orders in $\hbar$ when the theory is expanded about appropriate classical solutions.  Said differently, at least in favorable settings we expect that violations of microcausality at finite $\hbar$ can be made non-perturbatively small\footnote{There, however,  is an interesting question of whether this can be done with a fixed observables, or whether one expects only a family of observables labeled by an integer $n$ such the violations of microcausality involving each observable are $O(\hbar^n)$.  This is related to the question of whether expressions like \eqref{eq:SingIntSmeared} can be connected to well-defined non-perturbative observables, or whether they are intrinsically perturbative.}.   Our arguments are similar to those made in \cite{Khavkine:2012jf}, which proposes a related notion of quantum microcausality formulated in terms of gauge-dependent fields. We hope to see this conjecture probed in the near future by building on \cite{Brunetti:2013maa}, or by extending the two-dimensional analysis of \cite{Gary:2006mw} to dilaton gravity theories with a physical spaceitme metric (i.e., without Weyl-invariance, see footnote \ref{Weyl}).  As noted in section \ref{IR}, in the very interesting asymptotically AdS and dS contexts IR issues require non-perturbative modifications of our observables that should violate this mircolocality.  But such corrections are typically suppressed by some large scale, and in particular by the classical recurrence time in the AdS context.

At finite $\hbar$ we no longer expect diffeomorphism-invariance to be spontaneously broken in generic states.  To take an extreme example, consider an ensemble of states with finite energy in an infinite universe.  Quantum fluctuations blur out observables of the form \eqref{eq:SingInt} or \eqref{eq:SingIntSmeared} over spacetime scales set by the gradients of the $Z$-fields.  But an infinite universe with finite energy must have small gradients over most of its volume, so in these regions our observables will average over large regions of spacetime.  They will thus display non-zero commutators between observables that a priori seem to have large spacelike separations.   An exploration of the extent to which such observables can be in localized in space and time was begun in \cite{Giddings:2007nu}, but it would be interesting to map out such properties more generally, and especially in semiclassical states that describe black holes.

We also provided some brief remarks on other issues concerning gravitational observables and black holes.  In particular, we noted that the our construction allows a disturbingly-large number of observables with (perhaps exponentially) small mutual commutators to be localized in bags of gold (see figure \ref{gold}), though these commutators grow exponentially with time as in \cite{Shenker:2013pqa}.  As a side remark, we mentioned that observables naturally associated with Hawking radiation from evaporating black holes do not display this exponential growth and are immune from the associated chaos.

\section*{Acknowledgements}
It is a pleasure to thank Steve Giddings, Jim Hartle, and Joe Polchinski for conversations concerning gravitational observables spanning many years.   I am also grateful to Stefan Hollands and Bob Wald for similar conversations regarding perturbative quantum field theory. I once more thank both Steve Giddings and Stefan Hollands, as well as Igor Khavkine, for their comments on the conjecture of section \ref{quantum} and help with references.  I am further indebted to Bryce DeWitt for his many lessons concerning gravitational observables and their algebra during my PhD studies, and in particular his focus on the locality features of observables of single-integral type.  I am similarly indebted to Claudio Bunster (then Teitelboim) for his emphasis on the gravitational Gauss laws during the same period.  More recently, I have enjoyed useful discussions with Ahmed Almheiri, Tom Banks, Will Donnelly, Xi Dong, Dan Harlow, Juan Maldacena, Mark Srendnicki, and Douglas Stanford, and the other members of the Kavli Institute for Theoretical Physics (KITP) program ``Quantum Gravity Foundations: UV to IR."
This work was supported by the National Science Foundation under grant numbers PHY12-05500 and PHY15-04541 and by
funds from the University of California. I acknowledge the KITP for their hospitality during critical stages of the project
where this work was further supported in part by National Science foundation grant number PHY11-25915.

\end{document}